# Strain engineering on graphene towards tunable and reversible hydrogenation

*Zhiping Xu [1,*] and Kun Xue [2]*

[1] Department of Civil and Environmental Engineering, Massachusetts Institute of Technology, Cambridge 02139, MA, USA

[2] Department of Engineering Mechanics, Tsinghua University, Beijing 100086, China

[*] Corresponding author, email: xuzp@mit.edu

**ABSTRACT**

Graphene is the extreme material for molecular sensory and hydrogen storage applications because of its two-dimensional geometry and unique structure-property relationship. In this *Letter*, hydrogenation of graphene is discussed in the extent of intercoupling between mechanical deformation and electronic configuration. Our first principles calculation reveals that the atomic structures, binding energies, mechanical and electronic properties of graphene are significantly modified by the hydrogenation and applied strain. Under an in-plane strain of 10 %, the binding energies of hydrogen on graphene can be improved by 53.89 % and 23.56 % in the *symmetric* and *anti-symmetric* phase respectively. Furthermore the instability of *symmetrically* bound hydrogen atoms under compression suggests a reversible storage approach of hydrogen. In the *anti-symmetric* phase, the binding of hydrogen breaks the sp$^2$ characteristic of graphene, which can be partly recovered at tensile strain. A charge density based analysis unveils the underline mechanisms. The results reported here offer a way not only to tune the binding of hydrogen on graphene in a controllable and reversible manner, but also to engineer the properties of graphene through a synergistic control through mechanical loads and hydrogen doping.

Graphene, the ultimately mono-atomically layered material, possesses fascinating properties such as massless Dirac fermions [1], abnormal quantum Hall effects [2], and extremely high stiffness and structural stability[3]. Besides of these intrinsic properties, recently there have been arising efforts in the chemistry of graphene, especially in the extent of molecular doping and gas sensing [4-6]. The notion of doping has been renovated in a non-covalent and reversible manner thanks to the co-existence chemical inertion of sp$^2$ carbon network and activity of their $\pi$ electrons [4,7]. Gas sensors have been proposed as the result of their significant electronic conductance changes at various chemical environment [5]. This remarkable structure-property relationship can also be used to engineer the graphene sheet through functionalization. Hydrogenation [78] and epoxidation [9] are two of the most noticeable approaches in this direction. These modifications of graphene structures are able to introduce remarkable changes in the structural and electronic properties of intrinsic graphene. Recently experimental studies have shown that through hydrogenation, a metal-semiconductor transition occurs [7]. More interestingly the metallic characteristic and other novel properties of intrinsic graphene sheet can be recovered through structural annealing later on [7]. The reversible engineering on graphene has great potentials in the design of functional nanomaterials and applications. Reversible hydrogen binding on graphene is also promising for the hydrogen storage industry, which provides high hydrogen storage weight ratio up to 7.7 wt %, to meet DOE's 2010 goal (6 wt %). Novel and flexible controls on the hydrogenation, especially on its reversibility are extremely critical for these applications. Further development on this concept relies upon our understanding on the mechanism of reversible hydrogenation process. In this Letter, we studied the mechanism of hydrogenation from a combinational point of view in both structural and electronic properties. Mechanical deformation is applied to tune the hydrogenation of graphene. We find that, the binding energy of hydrogen on graphene depends strongly on the strain applied and could be changed by 50 % at a tensile strain of 10 %. At specific condition, these external mechanical controls can even introduce binding-releasing transition for hydrogen atoms. These results imply a new dimension for the precise control and design on the physical properties of graphene materials.

The hydrogenation of the graphene sheet can be synthesized experimentally in a well periodic manner [7]. There are two possible configurations for hydrogen binding. When the graphene monolayer is deposited or patterned on substrates, such as $SiO_2$ [10] or metals [11], only one side of the graphene can be hydrogenated (Figure 1(a)). The structure is called *symmetric* here or boat-like in Ref. [12]. On the other hand, if the graphene sheet is suspended partly on supports [13], the hydrogen atoms can be hosted on both sides of the graphene sheets (Figure 1(b)), i.e. in an *anti-symmetric* or chair-like phase. Due to the cooperativity of hydrogen binding, the hydrogen atoms tend to binding in cluster on the graphene sheet, to lower their total energy [14], The extremely hydrogenated graphene with a C:H ratio of 1:1 is called graphane [12,15], and will be investigated here.

The structures and properties of the graphene and graphane are investigated through first principle calculations here. In our work, plane-wave basis sets based density functional theory (DFT) was employed with local (spin) density approximation (LDA). We used the PWSCF code [16] for the calculation. The pseudopotential parameters were taken from Perdew-Zunger sets [17]. For all the results presented in this paper, energy cut-offs of 30 Rydberg and 240 Rydberg are used for plane-wave basis sets and charge density grids respectively. The settings have been verified to achieve a total energy convergence less than $10^{-5}$ Ry/atom. For variable-cell relaxation, the criteria for stress and force on atoms were set to be 0.01 GPa and 0.001 Ry/Å. Sixteen Monkhorst-Pack k-points were used in each periodic direction for Brillouin zone integration, which is qualified for the energy convergence criteria of 3 meV.

The structures of *symmetric* and *anti-symmetric* graphane obtained from variable-cell geometric optimization are shown in Figure 1. In our calculation, a unit cell of graphane containing two carbon and two hydrogen atoms are used to represent crystalline graphene, with the in-plane lattice constant *a*. The *anti-symmetric* phase is found to be 1.8 eV (per carbon-hydrogen pair) more energetically favorable than the *symmetric* one. The hydrogen atoms bound at the two inequivalent carbon atoms in graphene lattice repulse each other and thus favors the configuration where adjacent carbon-hydrogen pairs resides at opposite surface of graphene sheet. Local structures of *symmetric* and *anti-symmetric* graphane shown in Figure 1(c) and 1(d) present the characteristics of $sp^2$ graphene and $sp^3$ diamond characteristics respectively. In the *symmetric* phase (Figure 1(c)), binding of hydrogen atoms expands the lattice constant underneath hexagonal graphene lattice by 14 %, but preserves the planar configuration due to either the periodic boundary condition as applied in the calculation here or strong adhesion from the substrate in experiments. When these restrains are relaxed or removed, for example, when the graphene is weakly bound to the substrate, hydrogenation on one side can cause out-of-plane bending and form tubular structures [18]. The expanded but planar configuration of *symmetric* graphane implies that the $sp^2$ characteristic in graphene is preserved. Localized density of states (LDOS) plots near the Fermi energy confirms this. In Figure 2(a), the total density of states are projected onto s and p atomic orbitals on one carbon atoms. $p_z$ ($l = 1$, $m = 1$) orbital is found to contribute most near the Fermi level and the binding of hydrogen opens an energy gap of 0.26 eV. However in the *anti-symmetric* phase, symmetry breaking between the two representative carbon (hydrogen) atoms in the unit cell changes the $sp^2$ hybridization and introduces $sp^3$ characteristics. The $sp^3$ hybridization induces out-of-plane corrugation of 0.45 Å. The new carbon-carbon bond length is 1.52 Å that is close to 1.54 Å in diamond structure. Also the bond angle $A_{C-C-C} = 111.6°$ and $A_{H-C-C} = 107.3°$ resemble the tetrahedral angle 109°. The *anti-symmetric* phase has a band gap of 3.35 eV that is still lower than that in diamond (5.48 eV).

The abrupt changes of lattice constants and planar structures of graphene after hydrogenation in both *symmetric* and *anti-symmetric* phase suggest possible control of the hydrogenation process and the properties of graphane through applying mechanical deformation. We thus propose here that the binding of hydrogen atoms on graphene to be engineered through introducing in-plane deformation, i.e. by applying *a prior* strain to the graphene unit cell and structural relaxation afterwards in the calculation.

To quantify the binding strength between the hydrogen atoms and graphene sheet in terms of single carbon-hydrogen pair, we define the formation energy of graphane as $E_f = E_{graphane} - (E_{graphene} + E_H)$, where $E_{graphane}$ and $E_{graphene}$ are the energy of graphane and pristine graphene respectively. The reference energy value for hydrogen $E_H$ is chosen as −13.014 eV from spin-polarized calculation of isolated hydrogen atom. In absence of strain in the graphene lattice, *symmetric* and *anti-symmetric* graphane phases have binding strength $E_f$ of -1.044 and -2.847 eV/atom.

The strain effects on the binding strength are first investigated by applying biaxial loading, i.e. homogeneous expansion or contraction of the graphene. The graphene structure has an optimized lattice constant $a_g$ = 2.44 Å (see the inset in Figure 3(a)), while the lattice constant of *symmetric* graphane is optimized to be expaneded $a_{sym}$ = 2.79 Å. We investigate the structural stability of graphane through navigating the stress in structures under loading. Starting from the optimized structure and increasing (or decreasing) the lattice constant $a$, the stress in graphane first rises up in amplitude and then breaks down after a critical value $a_c$, after which the structure is considered to be mechanically unstable. In this definition, the *symmetric* phase is only stable with a between 2.6 and 3.4 Å. The structure fails with $a$ lager than 3.4 Å, where the sp$^2$ bond is broken. The energy of both graphene (with additional terms from isolated hydrogen atoms) and *symmetric* graphane increases as the tensile strain is enhanced, When the lattice constant $a$ is expanded to 3.35 Å, the binding energy of *symmetric* graphane exceed the value of undeformed graphene, as shown in Figure 3(a). Also we can see from the results that the energy of bare graphene increases much faster than *symmetric* graphane. As a result, the binding energy of hydrogen is changed by 53.89 % (with respect to the value without mechanical loads applied) when the graphene is under a pre-strain of 10 % (from $a$ = 2.44 to 2.69 Å). On the other hand, when the graphene sheet is restrained against to in-plane expansion or the *symmetric* graphane is compressed ($a$ lies below 2.65 Å), the repulsion between the carbon-hydrogen pairs causes the structure mechanically unstable. The hydrogen is repelled from the graphene plane (Figure 4). As a result, the hydrogen atoms will be released from graphene. The sensitivity of the *symmetric* hydrogenation of graphene on strain observed here suggests a flexible controllability by simply straining the graphene sheet. Moreover, a compressive loading on the graphane sheet is expected to be able to release hydrogen atoms and achieve reversible hydrogen storage. However, the ultra-thin sheet of graphene or graphane under compressive loads is supposed to buckle at a small critical strain. Thus to achieve the compression induced hydrogen releasing, the graphene should be deposited on a cohesive substrate or constrained in the *c*-axis direction in composites to prevent the in-plain strain energy release due to out-of-plane buckling.

In *anti-symmetric* phase, the optimized lattice constant $a_{asym}$ = 2.5 Å. A dependence of the formation energy on the strain is also found (Figure 3(a)). In-plane tensile and compressive strain of graphene sheet up to 10 % can change the binding strength remarkably by 23.56 and –2.9 % respectively. The structure breaks at a tensile strain of 22 % or a compressive strain of 26 %. In comparison with the *symmetric* phase, the strain modification of hydrogen binding energy is much smoother here. Under tensile strain, because of the decoupling between adjacent carbon-hydrogen pairs, the expansion of in-plane lattice constant is found to be able to reduce its out-of-plane corrugation slightly and partly recover the $\pi$ electron characteristics of graphene. Figure 4(a) shows the changes of carbon-hydrogen bond lengths $l_{C\text{-}H}$ in these two phases. In *symmetric* phase, $l_{C\text{-}H}$ increases at either tensile or compressive strain. As *a* is compressed to be less than 2.6 Å, the carbon-hydrogen bond is broken and hydrogen atoms are released from graphene sheet as discussed before. While at tensile strain and when the stress approaches the tensile strength $\sigma_s$ = 84.38 GPa and the in-plane sp$^2$ carbon network is about to be broken, $l_{C\text{-}H}$ is elongated to the same value as in methane ($l_{CH\text{-}methane}$ = 1.096 Å). On the other hand, the carbon-hydrogen bond length maximizes at the optimized configuration of *anti-symmetric* graphane. When tensile strain is applied, the planar configuration of graphene is partly recovered in the *anti-symmetric* phase. The carbon-hydrogen bond length also approaches $l_{CH\text{-}methane}$ as strain is increased. As shown in Figure 4(b), at an elongated lattice constant *a* = 3.05 Å, the charge density distribution of *anti-symmetric* phase of graphane resembles that of *symmetric* phases, which is also illustrated in Figure 4(c) that shows the rising significance of p$_z$ orbital, which shifts towards the Fermi level. This sp$^3$-sp$^2$ transition in the *anti-symmetric* phase shows that, as the distance between adjacent carbon-hydrogen pairs is increases, the coupling between the pairs that results in the sp$^2$-sp$^3$ transition from graphene to graphane is weakened.

The hydrogenation as a chemical functionalization of graphene not only modifies its electronic structures, but also affects its mechanical properties, as a result of the breaking of sp$^2$ network. Our results have shown that the in-plane modulus of graphene $C = d^2E/Ad\varepsilon^2$ = 1260 GPa has been reduced by 52 % and 26 % in *symmetric* and *anti-symmetric* phase respectively, where $E$ is potential energy, $\varepsilon$ is in-plane biaxial strain and $A$ is the calculated cross section area where the thickness of graphene is taken as 3.4 Å. Accordingly, the tensile strength decreases from 101.27 GPa to 49.64 and 67.92 GPa. The response of graphane structure under uniaxial loading is also investigated by applying tension in one direction and relaxing the unit cell length in the other direction, as shown in Figure 3(b). Similar effects of strain modulation on the hydrogenation binding energy are observed, although the change is less significant because of the transverse relaxation. The Young's modulus of graphene sheets $Y$ = 1060 GPa are found to be decreases by 77 % and 25 % in the *symmetric* and *anti-symmetric* phases.

In conclusion, we investigated the effects of hydrogenation on the structural, mechanical and electronic properties graphene. The focus is placed on the strain effects on the binding strength, in both *symmetric* and *anti-symmetric* phases. We found that mechanical deformation has remarkable effects on their binding strength: an in-plane strain of 10 % can induce up to 53.89 % change in the binding strength. More prominently, in-plane constraints and compressive strain on the graphene can cause its resistance to the hydrogen atoms or even releasing hydrogen atoms from *symmetric* graphane. The binding of hydrogen on graphene opens energy gap, reduces stiffness and strength of graphene remarkably. The observations here suggest novel approaches for both graphene materials engineering and hydrogen storage applications. Our work laid the ground for the design and application of functionalized graphene materials and functional devices such as chemical sensors through hydrogenation. The understanding of the underline structural-property relation thus obtained is also critically important for hydrogen storage applications with tunable and reversible controls from structural deformations and molecular doping synergistically.

**FIGURES AND CAPTIONS**

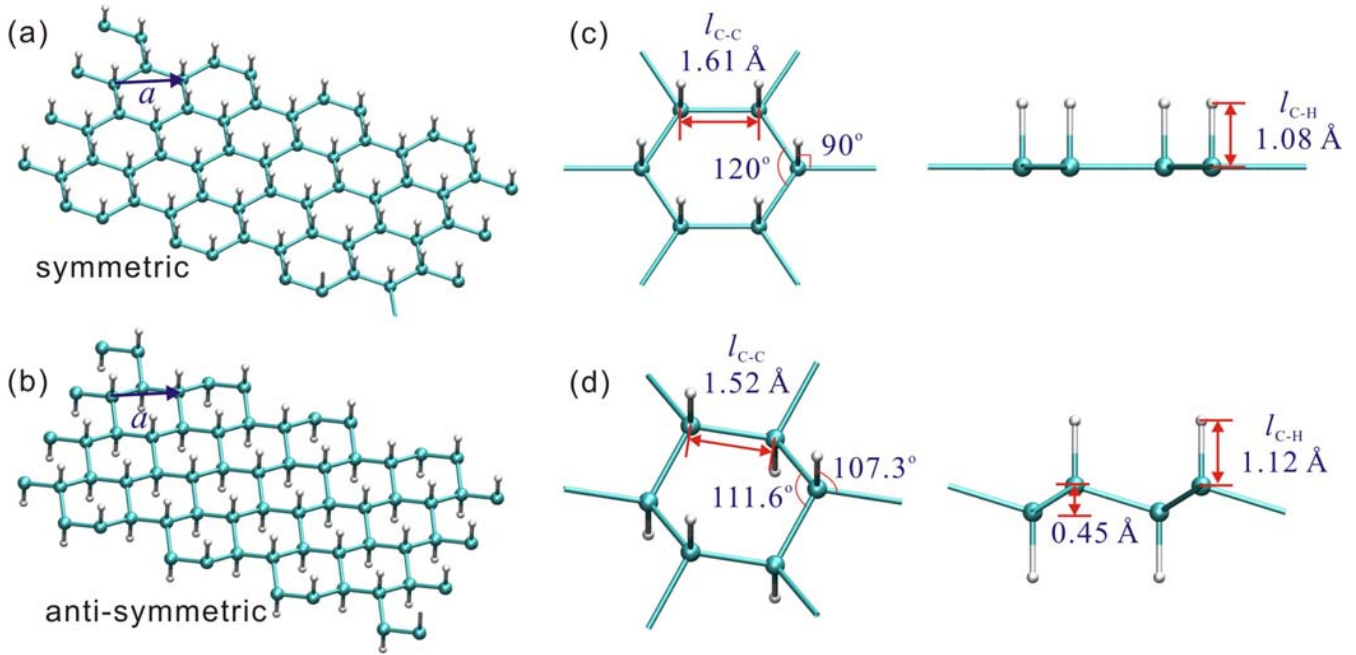

**Figure 1.** (a) and (b) Optimized atomic structures of *symmetric* and *anti-symmetric* graphane. The unit cell (marked by red box) contains two carbon atoms (in gray) as in graphene and two hydrogen atoms (in white). The *anti-symmetric* phase with hydrogen atoms arrange on opposite sides of the graphene (also known as chair-like [12]) is 1.8 eV (per carbon-hydrogen pair) more energetically favorable than the *symmetric* one where hydrogen atoms reside on the same side (also known as boat-like [12]). Panel (c) shows the local structure of *symmetric* phase. In the *symmetric* phase, binding of hydrogen atoms expands the lattice constant underneath hexagonal graphene lattice by 14%, but remains the planar configuration unchanged. (d) shows the local structure of *anti-symmetric* phase. The symmetry breaking modified the $sp^2$ hybridization by introducing $sp^3$ characteristics: the new C-C bond length is 1.52 Å, close to 1.54 Å diamond structure, and the bond angle C-C-C = 111.6° and H-C-C = 107.3° are close to the tetrahedral angle 109°. The $sp^3$ hybridization induces out-of-plane corrugation of 0.45 Å.

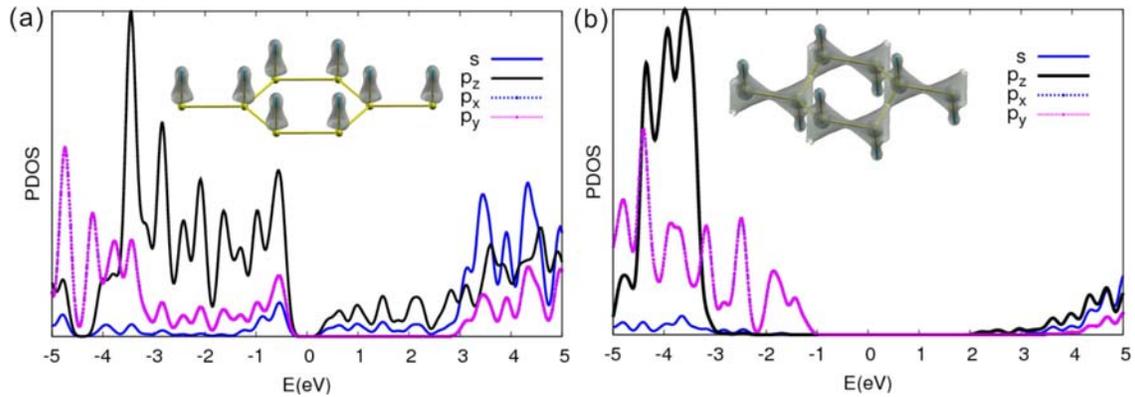

**Figure 2.** Density of states of graphane and its projection onto s, $p_x$, $p_y$ and $p_z$ orbitals of carbon atoms. In *symmetric* phase (a), the $p_z$ state (with angular momentum l = 1, m = 1) dominates the contribution to total density of states near the Fermi energy (referred as 0 eV). The hydrogenation opens a gap of 0.26 eV. The charge density plot shown as inset shows the $\pi$ characteristic in the expanded graphene sheets. In *anti-symmetric* phase (b), the sp$^3$ removes the $\pi$ characteristic and creates an energy gap of 3.35 eV instead. The inset shows sp$^3$ characteristic of the charge density in *anti-symmetric* graphane, which resembles the sp$^3$ bonds in diamond.

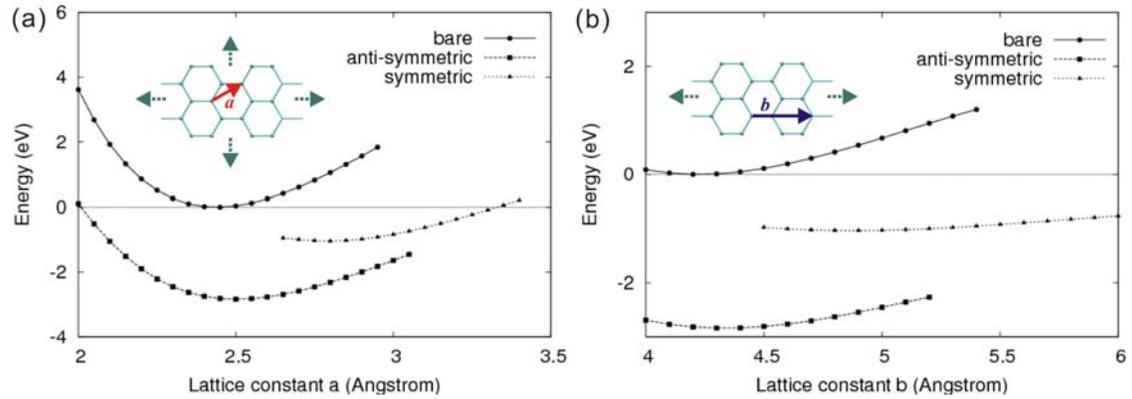

**Figure 3.** (a) Comparison between energies of graphene (with additional energy from isolated hydrogen atoms for comparison to the bare graphene) and graphane structures under biaxial strain loading (illustrated in inset). Only energy points for mechanically stable structures are plotted. The *symmetric* graphane phase has a larger lattice constant 2.79 Å as optimized than that of *anti-symmetric* graphane (2.50 Å) and graphene (2.44 Å). In their optimized configurations, the *symmetric* and *anti-symmetric* graphane have binding energies $E_b = (E_{graphane} - E_{graphene} + E_H)$ to hydrogen atom of –1.04 and –2.84 eV respectively. (b) Energies of graphene and graphane structures under uniaxial strain loading (as illustrated in inset).

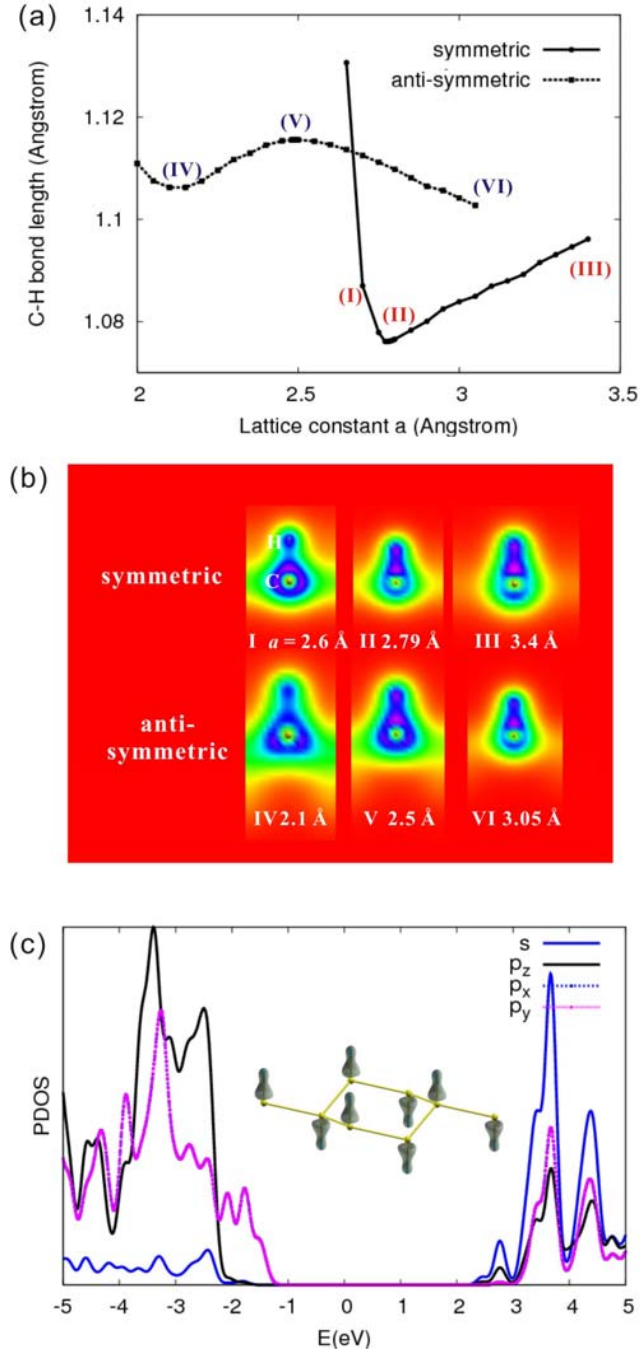

**Figure 4.** (a) Change of C-H bond length $l_{CH}$ under biaxial strain deformation. Six data points as denoted are selected for charge density distribution in Panel (b). In the *symmetric* phase, the carbon-hydrogen bond length under either tensile or compressive strain is elongated from its optimized value 1.08 Å. (c) shows the density of states of anti-symmetric graphane at an elongated lattice constant $a =$

3.05 Å. The $\pi$ characteristic of graphene is partly recovered because of the decoupling between adjacent carbon-hydrogen pairs at increased distance. The inset shows the isolevel of charge distribution. Charge density at $a = 2.65$ (I) shows the repelling of hydrogen from binding with graphene, in comparison with the optimized (II) and stretched (III) structures. For *anti-symmetric* phase, the carbon-hydrogen bond length maximizes at optimized lattice constant. The charge density doesn't change too much under compression (IV) in comparison with the optimized structure (V), but at tensile strain (VI) the out-of-plane corrugation of lattice is reduced slightly from 0.45 Å to 0.42 Å and some of the $sp^2$ and $\pi$ characteristics have been recovered, which is shown in the resemblance between charge densities at II and VI.

# REFERENCES

1. Novoselov, K. S.; Geim, A. K.; Morozov, S. V.; Jiang, D.; Katsnelson, M. I.; Grigorieva, I. V.; Dubonos, S. V.; Firsov, A. A., Two-dimensional gas of massless Dirac fermions in graphene. *Nature* **2005,** 438, (7065), 197-200.

2. Novoselov, K. S.; Jiang, Z.; Zhang, Y.; Morozov, S. V.; Stormer, H. L.; Zeitler, U.; Maan, J. C.; Boebinger, G. S.; Kim, P.; Geim, A. K., Room-Temperature Quantum Hall Effect in Graphene. *Science* **2007,** 315, (1379), 1137201.

3. Lee, C.; Wei, X.; Kysar, J. W.; Hone, J., Measurement of the Elastic Properties and Intrinsic Strength of Monolayer Graphene. *Science* **2008,** 321, (5887), 385-388.

4. Wehling, T. O.; Novoselov, K. S.; Morozov, S. V.; Vdovin, E. E.; Katsnelson, M. I.; Geim, A. K.; Lichtenstein, A. I., Molecular Doping of Graphene. *Nano Letters* **2008,** 8, (1), 173-177.

5. Schedin, F.; Geim, A. K.; Morozov, S. V.; Hill, E. W.; Blake, P.; Katsnelson, M. I.; Novoselov, K. S., Detection of individual gas molecules adsorbed on graphene. *Nat Mater* **2007,** 6, (9), 652-655.

6. Dan, Y.; Lu, Y.; Kybert, N. J.; Luo, Z.; Johnson, A. T. C., Intrinsic Response of Graphene Vapor Sensors. *Nano Letters* **2009,** 9, (4), 1472-1475.

7. Elias, D. C.; Nair, R. R.; Mohiuddin, T. M. G.; Morozov, S. V.; Blake, P.; Halsall, M. P.; Ferrari, A. C.; Boukhvalov, D. W.; Katsnelson, M. I.; Geim, A. K.; Novoselov, K. S., Control of graphene's properties by reversible hydrogenation: evidence for graphene. *Science* **2009,** 323, (5914), 610-613.

8. Singh, A. K.; Yakobson, B. I., Electronics and Magnetism of Patterned Graphene Nanoroads. *Nano Letters* **2009,** 9, (4), 1540-1543.